\documentclass[twocolumn,showpacs,preprintnumbers,amsmath,amssymb,prl]{revtex4-1}

\usepackage{graphicx}
\usepackage{dcolumn}
\usepackage{bm}
\usepackage{amsmath}
\usepackage{amssymb}
\usepackage{color}
\usepackage{braket}
\usepackage{lipsum}
\usepackage[pdffitwindow=true,colorlinks=true,frenchlinks=false,linkcolor=blue,anchorcolor=blue,citecolor=blue,filecolor=blue,
        urlcolor=blue,bookmarks=true,bookmarksopen=true,bookmarksnumbered=true,bookmarksopenlevel=1,plainpages=false,pdfpagelayout=TwoPageLeft,pdfpagelabels=true,breaklinks]{hyperref}

\begin{document}

\newcommand{\note}[1]{\textbf{#1}}

\title{Finite-time quantum entanglement in propagating squeezed microwaves}
\author{\firstname{Kirill G.} \surname{Fedorov$^{1}$}}
\email{kirill.fedorov@wmi.badw.de}

\author{\firstname{S.}~\surname{Pogorzalek$^{1,2}$}}
\author{\firstname{U.}~\surname{Las Heras$^{3}$}}
\author{\firstname{M.}~\surname{Sanz$^{3}$}}
\author{\firstname{P.}~\surname{Yard$^{1,2}$}}
\author{\firstname{P.}~\surname{Eder$^{1,2,4}$}}
\author{\firstname{M.}~\surname{Fischer$^{1,2,4}$}}
\author{\firstname{J.}~\surname{Goetz$^{1,2}$}}
\author{\firstname{E.}~\surname{Xie$^{1,2,4}$}}
\author{\firstname{K.}~\surname{Inomata$^{6,7}$}}
\author{\firstname{Y.}~\surname{Nakamura$^{6,8}$}}
\author{\firstname{R.}~\surname{Di Candia$^{9}$}}
\author{\firstname{E.}~\surname{Solano$^{2,3,5}$}}
\author{\firstname{A.}~\surname{Marx$^{1}$}}
\author{\firstname{F.}~\surname{Deppe$^{1,2,4}$}}
\author{\firstname{R.}~\surname{Gross$^{1,2,4}$}}
\email{rudolf.gross@wmi.badw.de}

\affiliation
{
$^{1}$ Walther-Mei{\ss}ner-Institut, Bayerische Akademie der Wissenschaften, D-85748 Garching, Germany \\
$^{2}$ Physik-Department, Technische Universit\"{a}t M\"{u}nchen, D-85748 Garching, Germany \\
$^{3}$ Department of Physical Chemistry, University of the Basque Country UPV/EHU, Apartado 644, E-48080 Bilbao, Spain \\
$^{4}$ Nanosystems Initiative Munich (NIM), Schellingstra{\ss}e 4, 80799 M\"{u}nchen, Germany \\
$^{5}$ IKERBASQUE, Basque Foundation for Science, Maria Diaz de Haro 3, 48013 Bilbao, Spain \\
$^{6}$ RIKEN Center for Emergent Matter Science (CEMS), Wako, Saitama 351-0198, Japan \\
$^{7}$ National Institute of Advanced Industrial Science and Technology, 1-1-1 Umezono, Tsukuba, Ibaraki 305-8563, Japan \\
$^{8}$ Research Center for Advanced Science and Technology (RCAST), The University of Tokyo, Meguro-ku, Tokyo 153-8904, Japan \\
$^{9}$ Freie Universit\"{a}t Berlin, Institut f\"{u}r Theoretische Physik, Arnimallee 14, 14195 Berlin, Germany
}

\date{\today}

\begin{abstract}
Two-mode squeezing is a fascinating example of quantum entanglement manifested in cross-correlations of incompatible observables between two subsystems. At the same time, these subsystems themselves may contain no quantum signatures in their self-correlations. These properties make two-mode squeezed (TMS) states an ideal resource for applications in quantum communication. Here, we generate propagating microwave TMS states by a beam splitter distributing single mode squeezing emitted from distinct Josephson parametric amplifiers along two output paths. We experimentally study the fundamental dephasing process of quantum cross-correlations in continuous-variable propagating TMS microwave states and accurately describe it with a theory model. In this way, we gain the insight into finite-time entanglement limits and predict high fidelities for benchmark quantum communication protocols such as remote state preparation and quantum teleportation.
\end{abstract}



\maketitle

Propagating quantum microwave signals in the form of squeezed states are natural candidates for quantum communication \cite{CVTele1998,QKeyD2012,CVarQI2005} and quantum information processing \cite{QCwCV1999,GausQI2012,CVQC2016,HybrQI2015} with continuous variables. This assessment stems from the fact that they belong to the same frequency range and are generated using the same material technology as quantum information processing platforms based on superconducting circuits. Utilizing propagating quantum microwaves, one can potentially realize quantum illumination protocols \cite{QuanIllu2008,ExpQIllu2015,MwQIllu2015,QICloak2016,QIPrinc2017}, hybrid computation schemes with continuous variables \cite{HybrQI2015,DispSMS2016,Qumode2016} and a high-fidelity seamless connection between distant superconducting quantum computers \cite{MwQTele2015}. In this context, finite-time correlation properties of propagating squeezed states provide necessary information about the tolerance to unwanted delays, and therefore, determine whether delay lines are required. From the fundamental point of view, finite-time correlation measurements grant a quantitative physical insight into the dephasing processes of propagating entangled microwave signals.

In this Letter, we experimentally show how the entanglement in the two-mode squeezed state decays depending on a time delay $\tau$ in one of the propagation paths. This entanglement decrease can be attributed to a dephasing process between the two modes propagating along each path. According to our experimental results and the corresponding theoretical model, the dephasing time, characterising the entanglement decay, is inversely proportional to the squeezing level and the measurement filter bandwidth.

We use two superconducting flux-driven Josephson parametric amplifiers (JPAs) operated at $f_0=5.323\,$GHz for the generation of squeezed microwave states \cite{SupMat}. The task of each JPA is to perform a squeezing operation on the incident vacuum state $\hat{S}(\xi) |0\rangle$, where $\hat{S}(\xi) \,{=}\, \exp(\frac{1}{2}\xi^* \hat{a}^2 \,{-}\, \frac{1}{2}\xi (\hat{a}^\dagger)^2)$ is the squeezing operator, $\hat{a}^{\dagger}(\tau)$ and $\hat{a}(\tau)$ are the creation and annihilation operators of the $f_0$ mode, and $\xi = r e^{i\phi}$ is the complex squeezing amplitude.  Here, the phase $\phi$ determines the squeezing angle in phase space, while the squeezing factor $r$ parameterizes the amount of squeezing. We conveniently characterize the degree of squeezing of the quantum state in decibels as $S\,{=}\,{-}10\,\log_{10} [\sigma_{\rm{s}}^2 / 0.25]$, where $\sigma_{\rm{s}}^2$ is the variance of the squeezed quadrature and the vacuum variance is $0.25$. Positive values of $S$ indicate squeezing below the vacuum level. In experiments, the JPAs are nonideal, and thus, add a finite number of noise photons $n$ to the squeezed vacuum. In this scenario, the squeezed and antisqueezed quadratures can be expressed as $\sigma_{\rm{s}}^2 = 0.25 (1+2n) \exp(-2r)$ and $\sigma_{\rm{a}}^2 = 0.25 (1+2n) \exp(2r)$, respectively \cite{SupMat}. This expressions allow us to rewrite the squeezing level as $S = -10 \log_{10} [(1+2 n) \exp( -2r)]$. Although $n$ is typically small, it must be accounted for in a quantitative analysis of actual data.

Finite-time correlations of a propagating single-mode squeezed state can be captured using the normalized second order correlation function $g^{(2)}(\tau) =  \langle \hat{a}^{\dagger}(0) \hat{a}^{\dagger}(\tau) \hat{a}(\tau) \hat{a}(0) \rangle / \langle \hat{a}^{\dagger}(0) \hat{a}(0) \rangle ^2$. It describes the decay of the correlations in propagating light and relates an autocorrelation time to this process.
\begin{figure}
        \begin{center}
        \includegraphics[width=\linewidth,angle=0,clip]{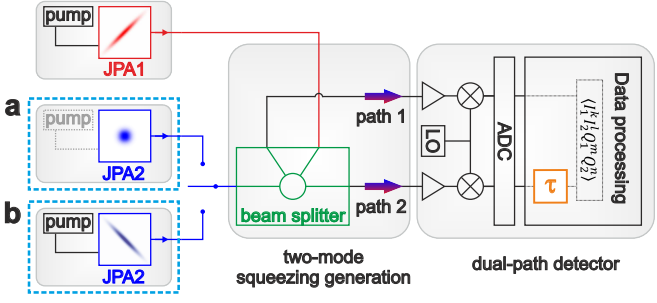}
        \end{center}
    \caption{Circuit schematic for the generation of squeezed propagating microwave states. In case \textbf{a}, JPA2 is in the vacuum state (the respective microwave pump is switched off) is depicted. In case \textbf{b}, both JPAs are operated such that symmetric TMS states are generated at the outputs of the hybrid ring. State detection, in both cases, is realized with the dual-path detector \cite{DPRecon2010,DPMethods2014} based on cross-correlation quadrature measurements. For finite-time correlation measurements, we introduce a time delay $\tau$ in one of the paths.}
   \label{fig1}
\end{figure}

To measure $g^{(2)}(\tau)$ of propagating squeezed microwaves, we employ the experimental setup shown in  Fig.\,\ref{fig1}a. In order to reconstruct the squeezed vacuum states, we apply a variant of the dual-path reconstruction scheme \cite{PathEnt2012,DPRecon2010,PhotStat2016}. To this end, the input signal is first distributed over two paths using a hybrid ring beam splitter. After amplification, auto- and cross-correlation measurements are performed on the output paths. In this way, we can correct for the amplifier noise and retrieve all moments of the signal mode  incident at the beam splitter up to fourth order in amplitude. An important modification in our setup is the introduction of a digital time-delay $\tau$ in one of the detection paths (see Fig.\,\ref{fig1}). This modification allows us to extend the dual-path technique to measure finite-time correlations $g^{(2)}(\tau)$. For further technical details, we refer the reader to the supplemental material \cite{SupMat}.
\begin{figure}
        \begin{center}
        \includegraphics[width=\linewidth,angle=0,clip]{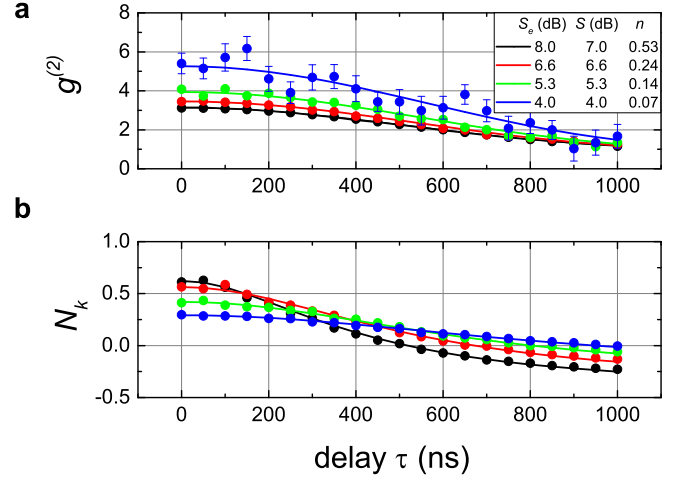}
        \end{center}
    \caption{Finite-time correlations of single-mode squeezed microwave states at the frequency $f_0=5.323\,$GHz. Top graph \textbf{a} illustrates the second order correlation function $g^{(2)}(\tau)$ of single-mode squeezed states fitted with equation (\ref{g2}). Bottom graph \textbf{b} depicts the impact of a finite-time delay on the path entanglement quantified via the negativity kernel $N_k(\tau)$, which can be accurately fitted using equation (\ref{Nk}). In both graphs, symbols depict experimental data, lines show corresponding theoretical fits, $S_e$ is the experimental squeezing level in JPA1, while $S$ is the fitted one, and $n$ is the fitted noise photon number. Fitting parameters and color codes are the same for \textbf{a} and \textbf{b}. When not shown, statistical error bars are smaller than the symbol size.}
   \label{fig2}
\end{figure}
\begin{figure*}
        \begin{center}
        \includegraphics[width=\linewidth,angle=0,clip]{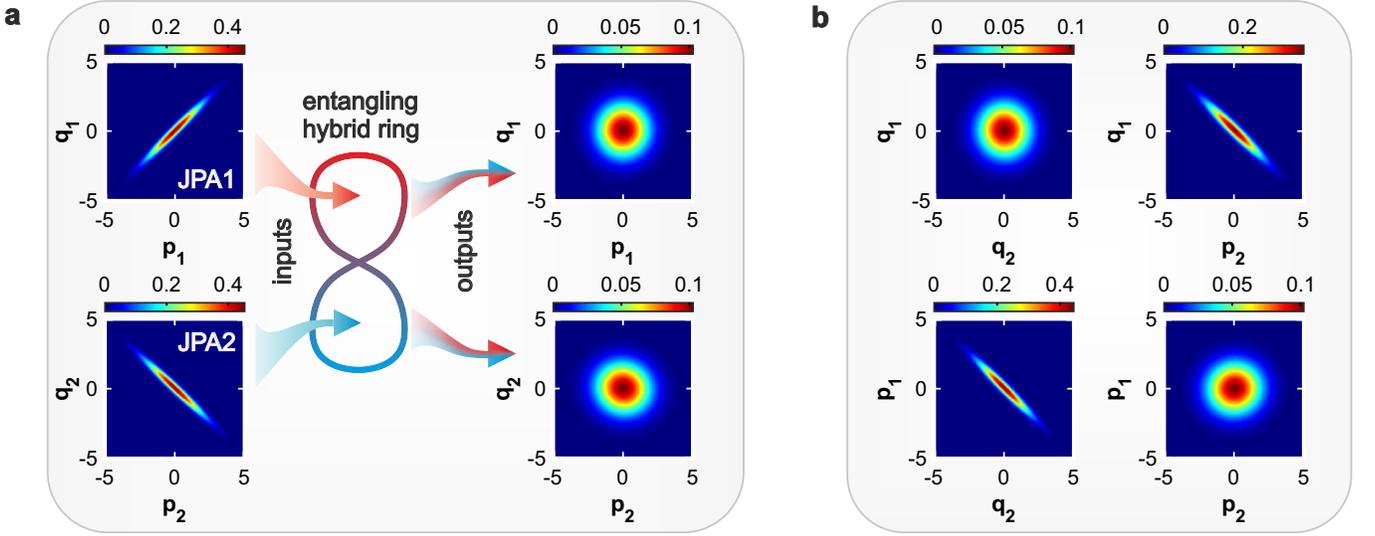}
        \end{center}
    \caption{Experimental marginal distributions of the Wigner function of TMS states created via interference of two orthogonally squeezed single-mode states with the squeezing levels $S_1 \simeq S_2 \simeq 8\,$dB in a hybrid ring. The quantities $p$ and $q$ are dimensionless variables representing quadratures of the electromagnetic field. Color reflects the quasiprobability amplitude. Left box \textbf{a} illustrates transformation of the hybrid ring inputs in the self-correlated subspaces $\{p_1, q_1\}$ and $\{p_2, q_2\}$. Its respective outputs show no quantum signatures and coincide with thermal states with an average number of photons $n_\mathrm{TMS} = 2.7$. Right box \textbf{b} depicts the output distributions in the cross-correlated subspaces $\{q_2,p_1\}$ and $\{p_2,q_1\}$ which uncover quantum entanglement in the form of two-mode squeezing with $S_\mathrm{TMS} = 7.2\,$dB and corresponds to the negativity kernel of $N_k = 2.1$.}
   \label{fig3}
\end{figure*}
Figure\,\ref{fig2} shows the experimental results. As expected from theory \cite{QOptics2007}, the super-Poissonian character of squeezed microwave states is demonstrated by $g^{(2)}(0) > 1$. We observe a smooth decay of $g^{(2)}(\tau)$ to the coherent-state limit $g^{(2)}(\tau) = 1$ on a timescale of $\tau \gtrsim 1\,\mu$s. In order to describe our results, we use an extended variant of the input-output theory presented in Ref.\,\onlinecite{g2Grosse2007}. It yeilds
\begin{equation}
\label{g2}
g^{(2)}(\tau) = 1 + {\rm sinc}^2(\Omega\tau)\frac{1+2\sigma_{\rm{s}}^2(\sigma_{\rm{s}}^2-1)+2\sigma_{\rm{a}}^2(\sigma_{\rm{a}}^2-1)}{(1-\sigma_{\rm{s}}^2-\sigma_{\rm{a}}^2)^2},
\end{equation}
where $\Omega$ is a width of a measurement bandpass filter centered at $f_0$. By fitting experimental data with equation (\ref{g2}), as depicted in Fig.\,\ref{fig2}a, we extract the experimental filter bandwidth $\Omega \simeq 420\,$kHz. This value agrees very well with the bandwidth of the digital finite-impulse response (FIR) filter used in our measurement protocol, $\Omega_{\rm FIR} = 430\,$kHz.  As can be seen from equation (\ref{g2}), the fitted squeezing factor $r$ and noise photon number $n$ define the value of $g^{(2)}(0)$, but neither one influences the temporal shape of $g^{(2)}(\tau)$.

So far, we have investigated the correlations of single-mode squeezed states. Now, in order to evaluate the suitability of our setup for actual quantum microwave communication protocols, we go one step further and study the finite-time behaviour of quantum entanglement between the outputs of the hybrid ring. In general, the amount of quantum entanglement can be assessed using the negativity $\mathcal{N}$. It is defined as $\mathcal{N} \,{\equiv}\, \max \left\{ 0, N_k \right\}$, where the negativity kernel $N_k$ is a function of the density matrix of the investigated state. In the relevant case of Gaussian states, $N_k$ is expressed in terms of the covariance matrix \cite{Negat2005,SupMat}, and $\mathcal{N}$ constitutes an entanglement measure. Then, condition $N_k > 0$ implies the presence of entanglement. By varying $\tau$ in our experiment, we measure the $\tau$-dependent negativity kernel $N_k(\tau)$ which provides information about the temporal length of the propagating entangled signal. Here, we introduce a maximally acceptable delay $\tau_d$ which still allows for existence of entanglement in a TMS state. This condition is given by $N_k(\tau_d) = 0$ for a monotonically decreasing $N_k(\tau)$. Assuming all other system properties to be equal, a large $\tau_d$ is beneficial in quantum communication. The expression for the negativity kernel $N_k(\tau)$ for an arbitrary two-mode squeezed state produced by two independent JPAs is given by \cite{SupMat}
\begin{equation}
\begin{split}
\label{Nk}
&N_{k}(\tau) = -0.5 + 0.5 \big[ (n_1 - n_2)^2 + \tilde{n} C + \\
&(\tilde{n} C - (n_1 + n_2 + 1)^2) \mathrm{sinc}^2\,\Omega\tau - \tilde{n} D \lvert\mathrm{sinc}\,\Omega\tau\rvert \big]^{-0.5},
\end{split}
\end{equation}
where $\tilde{n} \equiv (1 + 2 n_1)(1 + 2 n_2)$, $C \equiv \cosh^2(r_1 + r_2)$, $D \equiv \sinh (2 r_1 + 2 r_2)$; $r_1$ ($r_2$) and $n_1$ ($n_2$) are the squeezing factor and number of noise photons of JPA1 (JPA2), respectively. For the scenario depicted in Fig.\,\ref{fig1}a, where JPA2 is off, we consider $r_2 = n_2 = 0$ and accurately fit the experimental data with equation (\ref{Nk}) as it can be seen in Fig.\,\ref{fig2}b. The negativity kernel $N_k$ depends on the squeezing level $S$ and the corresponding values of $r$ since the delay $\tau_d$ clearly decreases with increasing $S$. This is in strong contrast with the behavior of $g^{(2)}(\tau)$ where the temporal shape only provides information on $\Omega$. The increasing noise photon numbers $n$ for higher squeezing levels are caused by larger microwave pump powers required to reach these values of $S$. As a result, stronger microwave fields lead to a stronger coupling to environmental loss channels, a process that increases $n$.

\begin{figure*}
        \begin{center}
        \includegraphics[width=\linewidth,angle=0,clip,bb = 0 20 648 380]{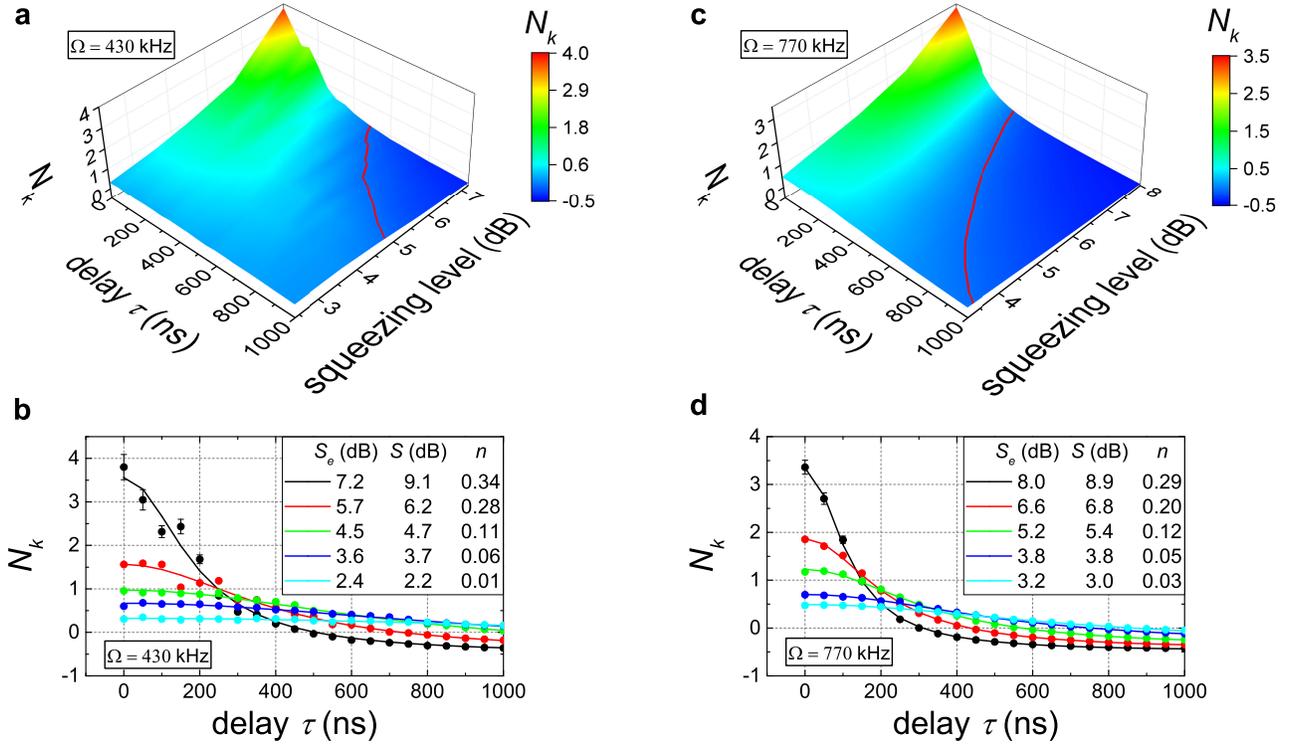}
        \end{center}
    \caption{Entanglement dephasing of propagating TMS states. The negativity kernel $N_k$ is plotted as a function of the delay time $\tau$. Left graphs (\textbf{a} and \textbf{b}) show experimental data measured with the filter bandwidth $\Omega = 430$ kHz. Right graphs (\textbf{c} and \textbf{d}) show the data obtained with $\Omega = 770$ kHz. The red line in \textbf{a} and \textbf{c} marks the boundary between the entangled ($N_k > 0$) and classical ($N_k < 0$, separable) regions. The circles in \textbf{b} and \textbf{d} represent the experimental data, the lines show the corresponding fits using equation (\ref{Nk}), $S_e$ values in the insets represent the experimental squeezing levels for both JPA1 and JPA2, $S$ values denote the fitted squeezing levels, and $n$ denotes the corresponding noise photon numbers. When not shown, statistical error bars are smaller than the symbol size.}
   \label{fig4}
\end{figure*}
Finally, we consider the case of propagating symmetric TMS states, which is most relevant for our envisioned applications. To this end, we operate both JPAs with orthogonal squeezing angles as shown in Fig.\,\ref{fig1}b. The resulting TMS states are essential for quantum microwave communication as they allow for entanglement distribution between separate distant parties. First, we are interested in the full tomography of the path-entangled two-mode states. We achieve this goal by using the reference state method \cite{PathEnt2012,DispSMS2016,DPMethods2014,TMS2011} and zero time delay $\tau = 0$. The result is depicted in Fig.\,\ref{fig3}. We tune the squeezing levels and angles of both JPAs by adjusting the amplitude and phase of the microwave pump tone to obtain symmetric TMS states. In this way, we eliminate residual single-mode squeezing in the marginal distributions in the self-correlated subspaces $\{p_1, q_1\}$ and $\{p_2, q_2\}$ of the resulting Wigner function (see Fig.\,\ref{fig3}a). We achieve close to perfect local thermal states in both paths. The strong two-mode squeezing between them is clearly observed in the cross-correlation subspaces $\{q_2,p_1\}$ and $\{p_2,q_1\}$ in Fig.\,\ref{fig3}b. This result is very different from the case shown in Fig.\,\ref{fig1}a, where, in addition to the two-mode squeezing, also a residual local single-mode squeezing is generated at the hybrid ring outputs. The latter is unwanted in certain communication protocols since it partially reveals the encoded information. To describe the entanglement for this case, we use equation (\ref{Nk}) in the approximation of identical squeezing factors $r_1 = r_2 = r$ (accordingly, $S_1 = S_2 = S$) and noise photon numbers $n_1 = n_2 = n$ in both JPAs. Figure\,\ref{fig4} describes the entanglement dephasing in the propagating TMS states with respect to the finite-time delay $\tau$ of the second path for various input squeezing levels and different filter bandwidths. The fit to the experimental data has been implemented using the squeezing factor and number of noise photons as free fitting parameters while setting the filter bandwidth $\Omega$ to the value predefined by FIR characteristics ($430$ kHz or $770$ kHz). Figure\,\ref{fig4} shows an excellent agreement between theory and experiment. Moreover, the entanglement dephasing time $\tau_d$ strongly decreases with the increasing squeezing levels $S$, and consequently, the squeezing factors $r$. Additionally, high squeezing levels lead to a substantial increase in the added noise photons $n$, which leads to weaker entanglement $N_k(\tau=0)$ but does not affect $\tau_d$. Both contributions may be seen in equation (\ref{Nk}). Another important experimental observation is that increasing the filter bandwidth $\Omega$ leads to shorter dephasing times $\tau_d$ which is also described by equation (\ref{Nk}). In the limit $\Omega \rightarrow \infty$, $\tau_d$ is eventually limited by the internal JPA bandwidth $\Delta f \simeq 5$ MHz.

The demonstrated dephasing process of TMS states allows us to estimate fidelities for two relevant quantum communication protocols based on propagating squeezed microwaves. The first one is remote state preparation (RSP) \cite{MwQTele2015}. It can be viewed as the creation of a remote quantum state with local operations. Here, we consider the case of $\Omega = 430$ kHz and $S_e \simeq 5.7$ dB. A straightforward estimation based on the observed $N_k(\tau)$ and otherwise ideal conditions provides the drop in fidelity from unity $F_\mathrm{RSP}(\tau\,{=}\,0) = 1$ to $F_\mathrm{RSP}(\tau) = 0.95$ for $\tau \simeq 90$ ns. The situation improves significantly when aiming at the remote preparation with lower squeezing level $S_e \simeq 3.8$ dB, where the same drop does not happen until $\tau \simeq 250$ ns. A similar analysis for the quantum teleportation (QT) protocol \cite{MwQTele2015} shows that the initial fidelities are lower (since it ultimately depends on the initial squeezing level and $F_\mathrm{QT} \rightarrow 1$ only for $S \rightarrow \infty$), but decrease slower with increasing $\tau$. For $S_e = 5.7$ dB, the change from $F_\mathrm{QT}(\tau\,{=}\,0) \simeq 0.80$ to $F_\mathrm{QT}(\tau) \simeq 0.75$ occurs in $\tau \simeq 380$ ns. These timescales make the implementation of RSP and QT with propagating squeezed microwaves possible without implementing delay lines, since the required delays due to feedback operations in these protocols are estimated to be around $\tau \simeq 100 - 200$ ns.

In conclusion, we have confirmed that the autocorrelation time of microwave single-mode squeezed states depends on the minimum filter bandwidth $\Omega$ in the measurement setup. In our specific setup, $\Omega$ is defined by the bandwidth of the digital FIR filter. We have uncovered that, for propagating microwave TMS states, the entanglement quantified via $N_k(\tau)$ survives for finite-time delays $\tau$. We express the negativity kernel $N_k(\tau)$ as a function of the squeezing level $S$ and the noise photon number $n$ in addition to $\Omega$. This function is given by equation (\ref{Nk}) and accurately describes the experimental results. The measurement bandwidth $\Omega$ also influences the timescale of entanglement decay due to dephasing in the case of propagating TMS states. High squeezing levels and accompanying large noise photon numbers additionally reduce the entanglement for finite delays between the paths. With respect to applications, a key result of this work is the observation of the characteristic dephasing timescale $\tau_d$ for two-mode entanglement which can be longer than $1$ $\mu$s for squeezing levels $S \simeq 3$ dB. This result confirms that propagating TMS microwave states are suitable for the remote preparation of squeezed states and short-to-medium distance quantum teleportation.

We acknowledge support by the German Research Foundation through FE 1564/1-1, Spanish MINECO/FEDER FIS2015-69983-P, UPV/EHU PhD grant, Basque Government Grant IT986-16, European Project AQuS (Project No. 640800), Elite Network of Bavaria through the program ExQM, the projects JST ERATO (Grant No. JPMJER1601) and JSPS KAKENHI (Grant No. 26220601 and  Grant No. 15K17731). E.S. acknowledges support from a TUM August-Wilhelm Scheer Visiting Professorship and hospitality of Walther-Mei{\ss}ner-Institut and TUM Institute for Advanced Study. We would like to thank K. Kusuyama for assistance with part of the JPA fabrication.

\end{document}